\documentclass{elsarticle}

\usepackage{amsmath}    
\usepackage{amsfonts}
\usepackage{epstopdf}

\usepackage{epsfig}

\newcommand{\ket}[1]{\left\vert#1\right\rangle}
\newcommand{\bra}[1]{\left\langle#1\right\vert}

\newcommand{\one}{\mbox{$1 \hspace{-1.0mm}  {\bf l}$}}
\newcommand{\ketbra}[2]{|#1\rangle\langle #2|}

\begin{document}

\title{Glued trees algorithm under phase damping}
\author[QUB1]{J. Lockhart}
\author[QUB2]{C. Di Franco}
\ead{c.difranco@qub.ac.uk}
\author[QUB2]{M. Paternostro}
\address[QUB1]{School of Electronics, Electrical Engineering and Computer Science, Queen's University, Belfast, BT7 1NN, United Kingdom}
\address[QUB2]{Centre for Theoretical Atomic, Molecular and Optical Physics, School of Mathematics and Physics, Queen's University, Belfast, BT7 1NN, United Kingdom}

\begin{abstract}
We study the behaviour of the glued trees algorithm described by Childs {\it et al.} in [STOC `03, Proc. 35$^{\rm th}$ ACM Symposium on Theory of Computing (2004) 59] under decoherence. We consider a discrete time reformulation of the continuous time quantum walk protocol and apply a phase damping channel to the coin state, investigating the effect of such a mechanism on the probability of the walker appearing on the target vertex of the graph. We pay particular attention to any potential advantage coming from the use of weak decoherence for the spreading of the walk across the glued trees graph.
\end{abstract}


\maketitle

\section{Introduction}    

One of the main difficulties for the grounding of a platform for quantum technologies is the effect of noise or ``decoherence'' on quantum states. No physical system is ever truly closed due to interactions with its environment. As a result of such interactions, the quantum state of the system will approach classicality, thus ceasing to be of interest for quantum-empowered protocols~\cite{NielsenChuang}. Before being able to create useful quantum technologies, we need to understand these processes, and eventually control them. 

An intriguing aspect of decoherence is that, in specific cases (such as quantum stochastic resonance~\cite{Gammaitoni}, to throw an example), weak decoherence mechanisms give rise to sizeable advantages in, say, the performance of some quantum protocols or the transport of excitations across a quantum medium. Such counterintuitive effects are tightly linked to quantum interference phenomena: decoherence changes the way the wave function of a given system evolves in time, thus affecting the occurrence of constructive and destructive interference. ``Accidental" constructive effects may be induced, without spoiling the working principle of a given quantum process, for sufficiently weak decoherence mechanisms.

All this is particularly important (and evident) in the quantum walk framework~\cite{Kempe}, whose advantage in terms of the spreading rate of the position of a walker on a given ``path" may be magnified by small degrees of phase noise. In this paper, we build on the already well established body of research into the behaviour of quantum walks when affected by decoherence, reported for instance in Refs.~\cite{Kendon,Alagic,fedichkin} and surveyed in Ref.~\cite{Kendondeco}.

After quickly revisiting the paradigm of quantum walks, we proceed to discuss the protocol under investigation, introducing phase noise and addressing the performance of the scheme for various strengths of such mechanism. 

\section{Quantum walks}
A quantum walk is best described as the quantum analogue of the classical random walk. However, unlike the classical random walk, the evolution of a quantum walk is entirely deterministic. Quantum walks of course allow for superposition states of the walker, enabling them to exhibit interesting behaviours not shown by their classical counterparts. A comprehensive survey of quantum walks, covering both continuous and discrete time variants, and detailing the behaviours of quantum walks on various structures, can be found in Ref.~\cite{Kempe}. In this paper, we focus on discrete time quantum walks.

A discrete time quantum walk operates within the Hilbert space $H = H_p \otimes H_c$, where $H_p$ -- known as the position space -- describes the position of the walker on a well-defined structure (here we shall refer to this structure as the walk's \emph{terrain}), and $H_c$ -- known as the coin space -- describes an additional degree of freedom affecting the evolution of the walk: this degree of freedom determines the walker's behaviour in the next time step. For the evolution of the walk, we define two operators: the shift operator, $S$, and the coin operator, $C$. The shift operator will ``move'' the walker on to a new part of its terrain, depending on the coin state. For example, if the terrain of a walk is a graph, and the walker is on some vertex of the graph, the shift operator will move it along one of the vertex's edges to another vertex. The coin operator is analogous to the flipping of a coin in a classical random walk, it will act on the coin space which in turn affects how the walk shall evolve in its terrain when the shift operator is applied. We move the walker along by one step by applying the coin operator followed by the shift operator; the state of the walker, starting in some initial state $\ket{\psi(0)}$, is thus described after $t$ total steps by
$$
\ket{\psi(t)} = [S(\one_p\otimes C)]^t \ket{\psi(0)},
$$
where $\one_{p}$ is the identity operator on the position space. In other words, we apply the coin operator and the shift operator $t$ times to the initial walker state.

In order to provide a complete view of the main features of the walk protocol, we now give some concrete examples of quantum walks.

\subsection{Discrete time quantum walk on a line}
To represent the walk terrain, a line, we shall use the set of integers. The walker can be anywhere on the line, so we give $H_p$ the basis $\{\ket{i} : i \in \mathbb{Z}\}$. As previously described, each step of the walk involves a coin flip and a shift. In the walk on the line, the walker has a ``choice'' of two directions, left and right. In the classical random walk, the decision of which direction to walk in at each step is reached by flipping a fair coin. Likewise, in our quantum walk we shall use a coin space of degree two, viz. $H_c$ is given the basis $\{\ket{0}, \ket{1}\}$.

We decide to use the Hadamard operator for our coin, as in Ref.~\cite{Kempe}. This has the effect of putting our walker into a superposition of coin states and will allow for interference to occur during the course of the evolution of the walk. With regards to the walker's behaviour on the terrain, the classical random walk will move one step to the left or one step to the right depending on the most recent coin flip. The same idea applies for the quantum walk. We define the shift operator as

$$
S \ket{p,c} = \begin{cases} \ket{p-1, c} , \mbox{if } c = 0, \\ 
\ket{p+1, c}, \mbox{if } c = 1. \end{cases} 
$$

Ambainis {\it et al.} have shown in Ref.~\cite{Ambainis} that the quantum walk on the line spreads out quadratically faster than the classical random walk on the line.

\subsection{Discrete time quantum walk on a k-regular graph}
\label{discreteongraph}

\begin{figure}[t]
  \centering
\hskip-0.5cm{\bf (a)}\hskip5.25cm{\bf (b)}
    \includegraphics[width=0.48\columnwidth]{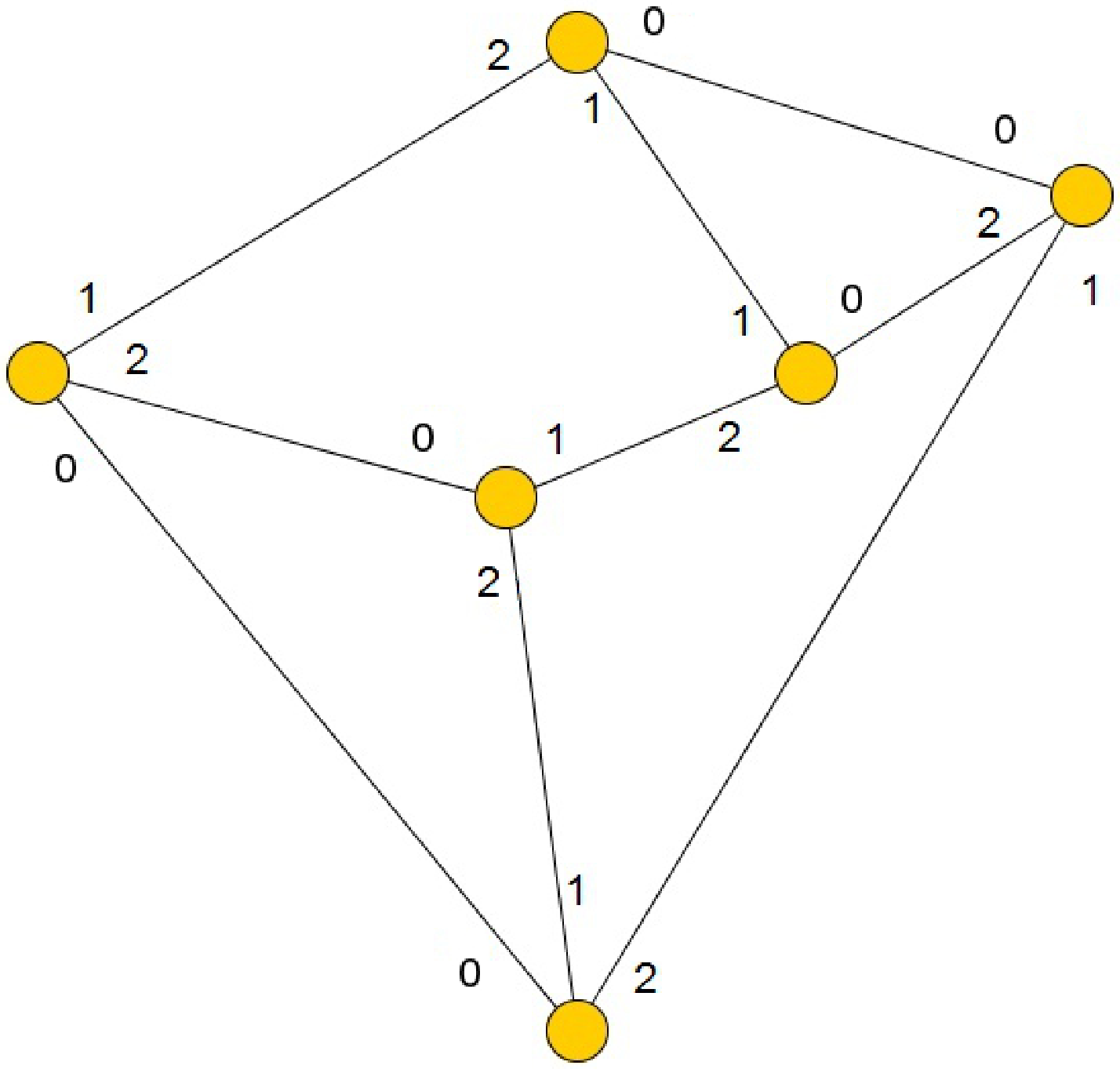}\includegraphics[width=0.52\columnwidth]{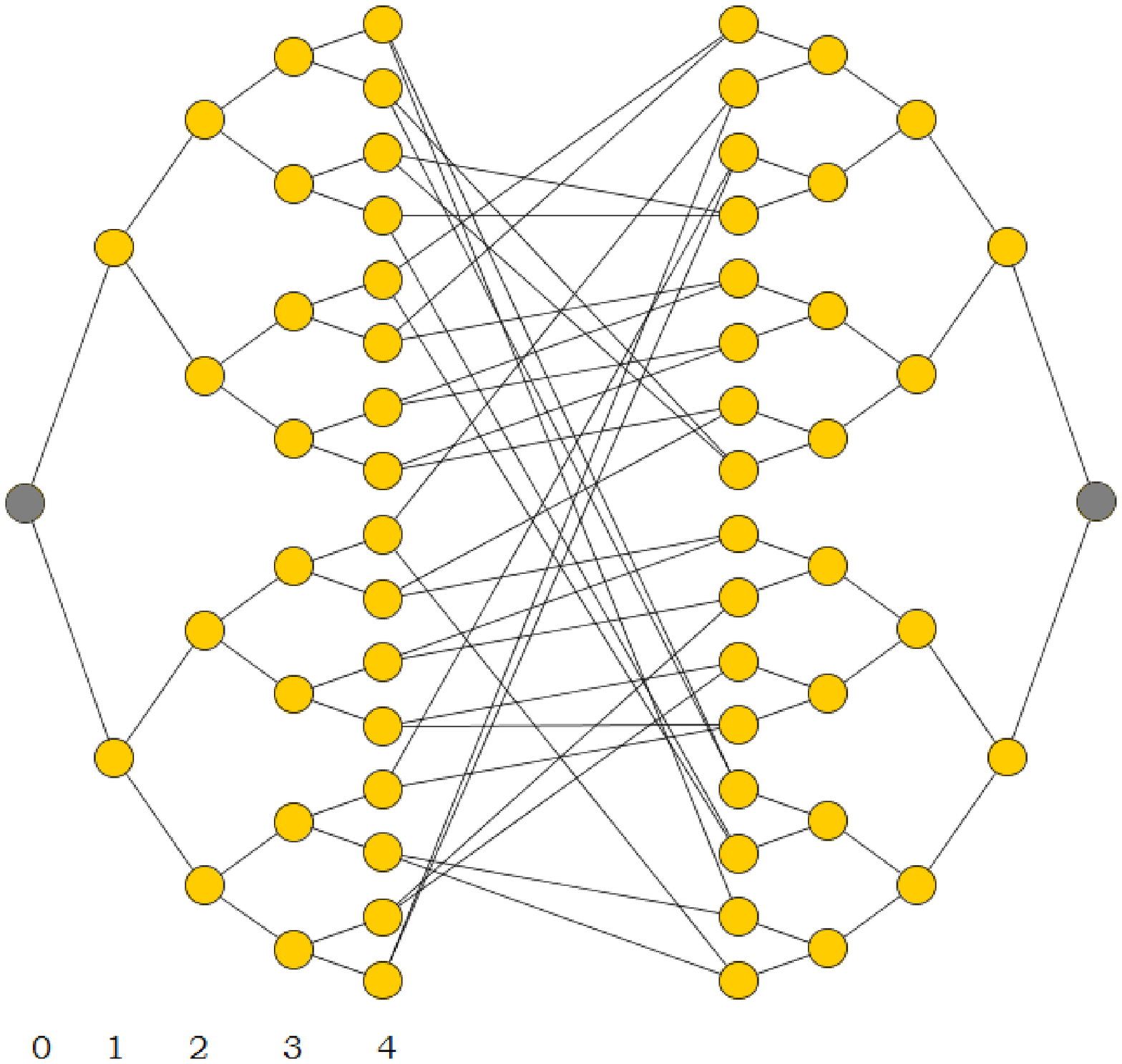}
  \caption{{\bf (a)} A labelled 3-regular graph. {\bf (b)} A glued trees graph with $4$ layers, $G'4$.}
  \label{labelled_graph}
\end{figure}

In general, a graph $G = (V,E)$ is specified by fixing a set of vertices $V$ along with a set of  edges $E$ connecting them. $k$-regular graphs are graphs with $k$ edges attached to each vertex. We affix a label $0 \le l \le k-1$ to each end of each edge, as illustrated in Fig.~\ref{labelled_graph} {\bf (a)}. The walker will traverse the graph's vertices, moving along the edges, so we define the position space $H_p$ as having the basis $\{\ket{p} : p \in V\}$.

At each time step, the walker has a fan-out of $k$ vertices to move to and we thus have to use an iso-dimensional coin space. We now give $H_c$ the basis $\{\ket{c} : 0 \le c \le k-1\}$. We then introduce the Grover coin  
\begin{equation}
C^{(G)}_{i,j} = \begin{cases} a, & \mbox{if } \delta_{i,j} = 1, \\
b, & \mbox{otherwise}, \end{cases}
\end{equation}
which, as described in Ref.~\cite{Kempe}, generalises the Hadamard coin to Hilbert spaces of dimension larger than 2. In order for $C^{(G)}_{i,j}$ to be unitary, the conditions $|a|^2+(k-1)|b|^2=1$ and $ab^*+a^*b+(k-2)|b|^2=0$ have to hold (with $a,b \in \mathbb{C}$). The values of $a$ and $b$ can be changed to vary the behaviour of the walk on the graph. We shall use a Grover coin later on to perform the simulations at the core of our work.

As for the shift operator, this must take the walker along the appropriate edge to a new vertex, depending on the coin state. Again, we state that this idea is a generalisation of the walk on the line in which we give the walker a choice of $k$ directions at each step rather than 2. We define our shift operator as
\begin{equation}
S\ket{v,c} = \ket{w,c'},
\end{equation}
where $(v,w)\in G$ and is labelled $c$ on $v$'s end, and $c'$ is the label assigned to the destination node's end of the edge.

\section{Model used}
The goal of the glued trees (GT) algorithm for quantum search is the following: beginning from the left-most vertex of a given GT graph, traverse the graph and reach the right-most vertex, referred to as the target vertex. Childs {\it et al.}~\cite{Childs} use this algorithm to show quantum walk search to be fundamentally more effective than classical random walk search by presenting a class of graphs (the GT graphs) that force classical random walks to make exponentially many queries to an oracle encoding the structure of the graph, but that are traversable by quantum walks with a polynomial number of queries to such an oracle.  In order to study the robustness of the algorithm to the detrimental effects of decoherence, we shall determine how effectively it achieves its goal when subjected to an increasing degree of phase damping noise. For this reason, we will focus on the probability that the walker is on the target vertex at the end of the walk. We thus consider GT graphs such as the one illustrated in Fig.~\ref{labelled_graph} {\bf (b)}, {i.e.} consisting of $n$ layers before the gluing stage, and thus labelled as $G'n$. 

The continuous time quantum walk exploited by Childs {\it et al.}~\cite{Childs} can be reformulated as a discrete time one by means of a shift operator analogous to the one that has been previously described, and a Grover coin with $a = -{1}/{3}$ and $b = {2}/{3} $ as described in Ref.~\cite{Tregenna}. More explicitly 
\begin{equation}
 C^{(G)} = \frac{1}{3}
 \begin{pmatrix}
  -1 & 2 & 2 \\
  2 & -1 & 2 \\
  2 & 2 & -1 \\
 \end{pmatrix}.
\end{equation}
We shall use this discrete time reformulation of the protocol to study its behaviour when affected by phase damping decoherence. The GT graphs described in Ref.~\cite{Childs} can straightforwardly be converted to $3$-regular graphs by adding two self loops (one each to the left-most and right-most vertices), thus allowing us to use the discrete time walk model described in Sec.~\ref{discreteongraph}. To model decoherence we use the Kraus operators for a phase damping channel acting on the $D$-dimensional system embodied by the coin only. This modifies the initial density matrix $\rho(0)$ of a system~\cite{NielsenChuang} as
\begin{equation}
\rho(\tau)=\mathcal{E}[\rho(0)] = \sum_{k}E_k(\tau) \rho(0)E^{\dagger}_{k}(\tau),
\end{equation}
where we have introduced the Kraus operators for phase damping~\cite{Liu,Pirandola}
\begin{equation}
E_k=\sum^{D-1}_{l=0}\frac{(l\sqrt{-2\ln\eta})^k\eta^{l^2}}{\sqrt{k!}}I_p\otimes\ket{l}\bra{l}_c
\end{equation}
with $\eta\in[0,1]$ the strength of the phase damping~\cite{Anosov}. By taking the parameterisation $\eta=e^{-\gamma \tau}$ with $\gamma$ the probability rate of a phase error, we can say that the effect of the channel is weak (strong) for $\eta\to1$ ($\eta\to0$). We have introduced the orthonormal basis $\{\ket{l}_c\}$ spanning the coin space only. We assume an initial position-coin state decomposed as
\begin{equation}
\label{ini}
\rho_{pc}(0)=\sum_{x,y,l,l'}\rho_{x,l,y,l'}\ket{x}\bra{y}_p\otimes\ket{l}\bra{l'}_c,
\end{equation}
where we have introduced the orthonormal basis $\{\ket{x}_p\}$ spanning the position space, and the position-coin density matrix elements $\rho_{x,l,y,l'}={}_{p,c}\!\bra{x,l}\rho_{pc}(0)\ket{y,l'}_{p,c}$. Eq.~\eqref{ini} evolves under the action of the phase damping channel on the coin space as
\begin{equation}
\label{evo}
\mathcal{E}'[\rho_{pc}(0)]=\sum_{x,y}\sum^{D-1}_{l,l'=0}\rho_{x,l,y,l'}\eta^{(l-l')^2}\ketbra{x}{y}_{p}\otimes\ketbra{l}{l'}_c.
\end{equation}
Our investigation consists of simulating a process such that, for each walk, we select a value of $\eta$, apply the channel to the density matrix of the system after each time step of the walk, and evaluate the probability of the walk reaching a given vertex of the graph. In this way, we compare the behaviour of the walks affected by phase damping ($0<\eta <1$) to that which is found for an ideal walk (corresponding to $\eta = 1.0$). It is important to note that the probability depends on the initial coin state. In our study we have considered walks starting in the initial coin state that maximises the performance of the ideal protocol (such state also depends on the way in which we label the graph; in particular, for the labelling scheme that we have mainly used, this corresponds to $\ket{\phi_0}=\alpha\ket{0}+\beta\ket{1}+\beta\ket{2}$, with $\beta\approx0.638$ and $\alpha=\sqrt{1-2\beta^2}$). However, we have tested different ways of labelling the graph (and therefore different initial coin states that maximise the walk's performances) and we have obtained results similar to those presented in the remainder of the paper.

\begin{figure}[t]
  \centering
    \includegraphics[width=0.625\columnwidth,angle=0]{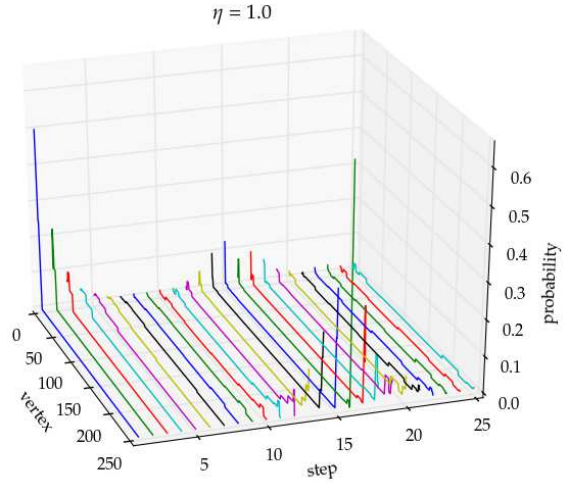}
  \caption{Probability of the walker being on specific vertices over time for the ideal walk (the walk unaffected by decoherence).}
  \label{nodeco}
\end{figure}

In order to grasp the temporal behaviour of the walk, we consider the change in the probability distribution of the walker's position on the graph at time $t$. In Fig.~\ref{nodeco} we illustrate such a probability distribution for an ideal 25-step walk on the GT graph $G'6$. The target vertex (vertex number 253) is reached, with the highest probability, on step 16 (as shown by the large peak in the line corresponding to step 16).

We will now focus on the probability of the walker being on particular vertices on the graph and the overall vertex probability distribution for specific time steps. In Ref.~\cite{Kendon}, the effect of decoherence on the walk on the hypercube was studied quantitatively. Starting from one corner, a decohered walker will reach the opposite corner in less time than in the ideal walk, hence highlighting a counterintuitive beneficial effect of such an incoherent process. Our goal here is to address similar questions for the walk on a GT graph, as well as to investigate the limits of validity of the claim made in Ref.~\cite{Kendon}, where a lingering effect of a decohered walk on the target vertex was suggested. We will thus look for the possibility that a decohered walk reaches the target vertex in less steps than the ideal one, and also attempt to determine whether it ``lingers'' on the target vertex for a longer time than in the ideal walk. 

\section{Discussion of results}
Here we present and discuss the results of our analysis, showing the behaviour of walks affected by decoherence of variable magnitude. All the results reported in this Section, unless otherwise specified, were generated by simulating the previously discussed discrete time reformulation of the algorithm by Childs {\it et al.}~\cite{Childs} on the GT graph $G'6$.

\begin{figure}[t]
  \centering
    \includegraphics[width=0.875\columnwidth,angle=0]{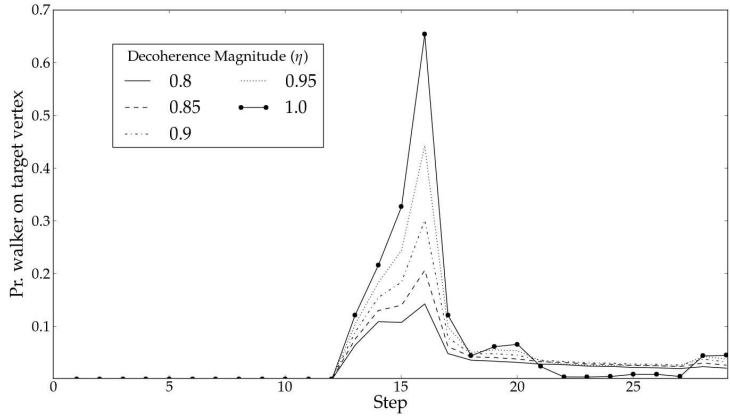}
  \caption{Probability of the walker being on target vertex over time for range of decoherence magnitudes.}
  \label{target_over_time}
\end{figure}

In Fig.~\ref{target_over_time} we plot how the probability of the walker being on the target vertex changes over time for various decoherence magnitudes, ranging from the ideal walk with $\eta=1.0$ to a decohered walk with $\eta=0.8$. We see from this plot that the ideal walk takes 13 steps to achieve its goal of reaching the target vertex, where we say that the walker has reached some vertex $v$ on some step $t$ of the walk when there is a non-negligible probability of the walker being on $v$ on the $t^{th}$ time step. The ideal walker first appears on the target vertex on step 13 with probability $P \approx 0.122$, then the probability comes to a peak on step 16 with a value $P \approx 0.655$ before steadily decreasing. We henceforth refer to the probability that a walker is on some vertex $v$ at a time step $t$ as $v$'s \emph{vertex probability} at time step $t$. We see in Fig.~\ref{target_over_time} that, as $\eta$ decreases, the target vertex probability steadily decreases. The peak on step 16 decreases at a faster rate than the target vertex probabilities associated with steps 13, 14, 15 and 17.

\begin{figure}[t]
  \centering
   \includegraphics[width=0.5\columnwidth]{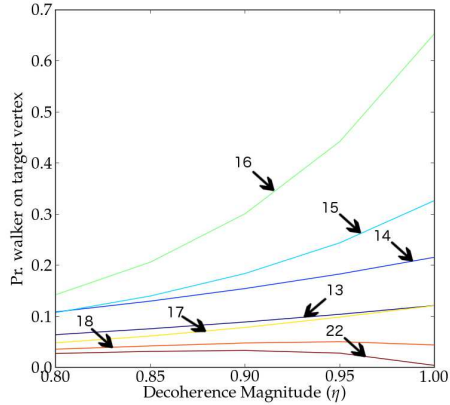}
   \caption{Probability of the walker being on target vertex over decoherence magnitude for range of step numbers.}
  \label{target_over_mag}
\end{figure}

Some interesting behaviour can be observed on steps 21 through 27 of the walk. Decoherence magnitude $\eta < 1.0$ slightly increases the target vertex probability; in other words, the ideal walker is less likely to be on the target vertex on steps 21 through 27 than the decohered walkers. This behaviour was reported in Ref.~\cite{Kendon} and can be seen more clearly in Fig.~\ref{target_over_mag}, in which we plot the effect of decoherence on the target vertex probability on various steps of the walk. In Fig.~\ref{target_over_mag}, in order to present this behaviour more clearly, we show a curve for step 22. We see, for step 22, that the target vertex probability is higher for walks affected by decoherence of magnitude $\eta < 1.0$, with the probability peaking at $\eta \approx 0.9$. Decoherence has spread out the range of time steps over which the target vertex probability is large compared to the classical
value. In line with the claims in Ref.~\cite{Kendon}, we have indeed found that weak decoherence increases the target vertex probability for an extended number of time steps. However, when we consider the difference between the $\eta=0.9$ and $\eta=1.0$ walks on step 22, we see that the difference between the two target probabilities is very small. In our simulations we have determined that there is a difference of approximately $0.0294$. As previously stated, in our simulations we have confirmed that this ``lingering'' effect lasts until step 27.

Fig.~\ref{target_over_mag}, as previously discussed, is a plot of the effect of decoherence on the target vertex probability on various time steps. The curve representing step 16 illustrates the effect of decoherence very well, and confirms the results reported in Ref.~\cite{Kendon}: as the decoherence magnitude $\eta$ decreases, the target vertex probability decreases exponentially. In other words, the algorithm by Childs {\it et al.}~\cite{Childs} becomes exponentially less effective at achieving its goal as $\eta$ decreases.

\begin{figure}[t]
  \centering
    \includegraphics[width=0.825\columnwidth,angle=0]{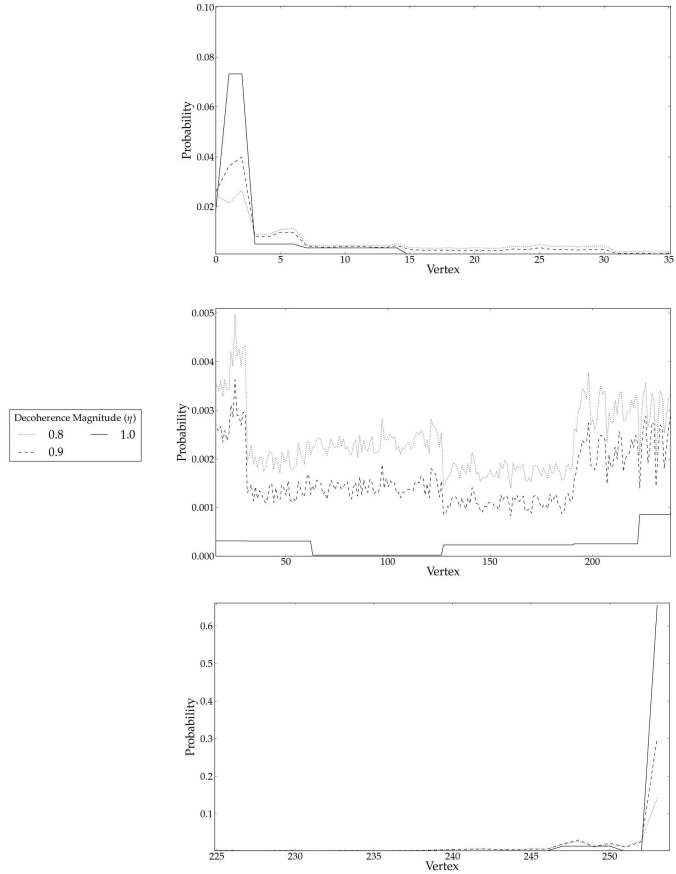}
  \caption{Change in vertex probabilities on step 16 for range of decoherence magnitudes. Note that each plot has different vertex numbers on the x-axis and different ranges of probability on the y-axis.}
  \label{step16zoom}
\end{figure}

We shall now investigate the extent of the ``damage'' that decoherence has on the algorithm's effectiveness. We concede that as long as the target vertex probability is higher than the other vertex probabilities then the decoherence has not had a particularly damaging effect on the effectiveness of the scheme. On the other hand, if phase damping decreased the peak associated with the target vertex below the other probability peaks then the algorithm would end up in a state involving a more probable ``false-positive'' than a ``positive'' when affected by decoherence -- we would say that this is a serious blow to any scheme's usefulness. In Fig.~\ref{step16zoom} we have plotted the entire graph's vertex probabilities on step 16 (the step on which, as previously mentioned, the probability of the ideal walker being on the target vertex was the highest) of walks with various decoherence magnitudes. Observe that the probability of the target vertex (corresponding to vertex number 253) is always higher than the non-target vertex probabilities, regardless of the value of $\eta$.

\begin{figure}[t]
  \centering
    \includegraphics[width=0.825\columnwidth,angle=0]{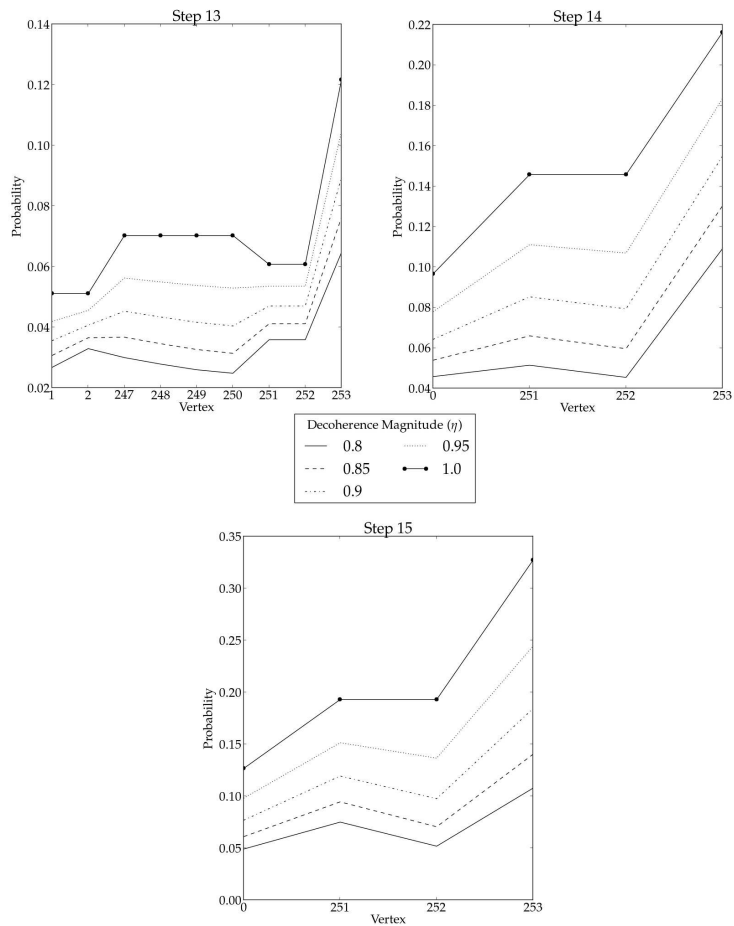}
  \caption{Target probability peaks compared to other peaks on steps 13 to 15.}
  \label{step13tophits}
\end{figure}

With Fig.~\ref{step13tophits} we can get a clearer picture on how the probability peaks change in the walks affected by decoherence. In order to improve the legibility of the plots while maintaining the number of shown vertex probabilities at an acceptable level, we present only the vertex probabilities $P > P_t/4$ where $P_t$ is the target vertex probability. From Fig.~\ref{step16zoom} and Fig.~\ref{step13tophits} we can see that, in steps 13 to 16 of the walk, the peak representing the target vertex probability never drops below any of the other vertex probability peaks. We find this significant and can conclude from it that the algorithm by Childs {\it et al.}~\cite{Childs} is not affected to the extent previously described: the walker is never on a non-target vertex with greater probability than the target vertex when allowed to run for sufficient time and phase damping decoherence does not change this fact.

The authors of Ref.~\cite{Kendon} investigate the evolution of a discrete time quantum walk on the hypercube using a Grover coin. They plot the probability that the walker is on the target vertex (they begin the walk on a corner of the hypercube and take the target vertex to be the vertex in the opposite corner of the hypercube) for a number of time steps, in the same way we have done in this paper. They observe that decoherence lowers the probability peaks of their plots in the same way that it does in our plots regarding the walk on the GT graphs, but they also observe that the ``troughs'' in their plots (sections of the curve representing vertices with probability much lower than others) become raised when decoherence is applied. This same effect can be observed in the walks that we simulated on the GT graphs and can be seen in the middle plot on Fig.~\ref{step16zoom}, where the vertices have their probabilities boosted slightly by the decoherence, with the lowest decoherence magnitude we investigated, $\eta=0.8$, raising the probability by the greatest amount.

\begin{figure}[t]
  \centering
   \includegraphics[width=0.625\columnwidth]{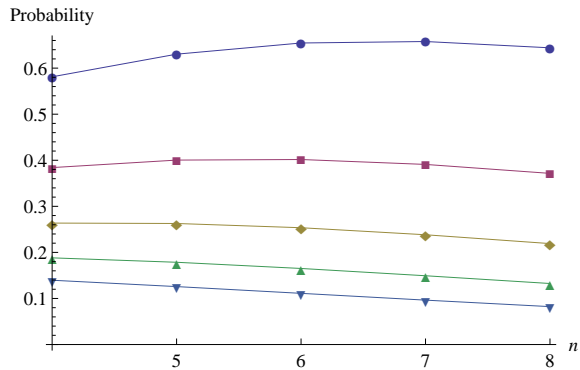}
  \caption{Probability of the walker being on target vertex (considering the step on which the probability is the highest) over the number $n$ of layers before the gluing stage in the GT graph, for range of decoherence magnitudes ($\eta=1, 0.95, 0.9, 0.85, 0.8$ from the top to the bottom, respectively).}
  \label{plotvslayers}
\end{figure}

For the sake of completeness, we finally extend our investigation to different sizes of the GT graph. We show in Fig.~\ref{plotvslayers} the behaviour of the target vertex probability against the number $n$ of layers before the gluing stage in the GT graph ({i.e.} we have generated GT graphs $G'n$, with $n$ between 4 and 8). We focus our interest on the step on which the probability of the walker being on the target vertex is the highest. Also in this case we consider a range of decoherence magnitudes $0.8 \le \eta \le 1.0$. The range of values for $n$ has been chosen as a reasonable trade-off between the computational power required by the simulation and the readability of the plot.

\section{Conclusions}
In this paper we discussed a discrete time reformulation of the continuous time quantum walk algorithm described by Childs {\it et al.} in Ref.~\cite{Childs}. We simulated this discrete time quantum walk on GT graphs and applied phase damping to the coin system, in order to analyse the algorithm's resilience to decoherence. We did this by studying how effectively it achieved its goal when affected by decoherence of various magnitudes: we investigated how decreasing the decoherence magnitude $\eta$ (making the decoherence ``stronger'') lowered the probability of the walker being on the target vertex at the end of the walk. We first simulated the walk with no decoherence to find how many steps it took to reach the target vertex (we say that the walk has reached a vertex when the probability that the walker is on that vertex is non-negligible). We then included a range of decoherence magnitudes $0.8 \le \eta < 1.0$ to see the extent of the drop in the target vertex probability at the end of the walk.

We observed that the ideal walk found the target vertex at time step 13, with a probability peak at step 16 (for a GT graph $G'6$). We noted that strengthening the decoherence (decreasing $\eta$) lowered the target vertex probability on time steps 13 to 17, but that the target vertex always had a higher probability than any other vertex, regardless of decoherence magnitude. We also observed that a decoherence magnitude of $\eta < 1.0$ boosted the target vertex probability very slightly above the ideal walk's ($\eta=1.0$) vertex probability on steps 21 through 27. Finally, we found that vertices with very low vertex probability had this boosted slightly by decoherence.

Our results have touched on some unexplored features of the algorithm by Childs {\it et al.}~\cite{Childs} on GT graphs, in turn opening up new questions to address. The first of such behaviours is the rate at which the target vertex probability on time step 16 decreases with $\eta$: out of the time steps 13-17, step 16's target vertex probability decreases at the highest rate. The second behaviour that deserves a deeper investigation in the future is observed in steps 21 through 27 of the walks studied in this paper. When compared to the target vertex probability for the ideal walk, we see a slight increase in the target vertex probability for $\eta < 1.0$ walks. To give a more specific example, we see an increase of $\approx 0.0294$ in the target vertex probability for $\eta=0.9$ on step 22 when we compare it to ideal walk. Finally, the boosting of the troughs in Fig.~\ref{step16zoom} by decoherence, which appears to be a similar phenomenon to the boosting of the target vertex probability in steps 21 through 27.

Our results add more weight to the claims made in Ref.~\cite{Kendon} on decohered walks: the ``lingering'' effect is shown to be present, but only marginally relevant because the boost in probability caused by the decoherence for the steps after the target vertex probability drop (step 18) is very small. On the other hand, we have observed that decoherence does not cause any upset to the notion that, at the end of the walk, the walker should be on the target vertex with a higher probability than any other vertex.

\section*{Acknowledgments}
This work has been supported by the UK EPSRC through a Career Acceleration Fellowship, a grant under the ``New Directions for Research Leader" initiative (EP/G004579/1), and the equipment grant (EP/K029371/1). JL thanks the Centre for Theoretical Atomic, Molecular, and Optical Physics for hospitality during the early stages of this work.

\end{document}